\documentclass[11pt,twoside]{article}
\usepackage{./asp2010}

\resetcounters

\usepackage{epsfig}
\begin{document}

\title{Strange mode instability for micro-variations in  Luminous Blue Variables}
\author{Hideyuki Saio$^1$, Cyril Georgy $^2$, and Georges Meynet$^3$
\affil{$^1$Astronomical Institute, Graduate School of Science, Tohoku University, Sendai, Japan}
\affil{$^2$Centre de Recherche Astrophysique de Lyon, \'Ecole Normale Sup\'erieure de Lyon, 46, all\'ee d$^,$Italie, 69384 Lyon Cedex 07, France; \\ 
Astrophysics, Lennard-Jones Laboratories, EPSAM, Keele University, ST5 5BG, Staffordshire, UK}
\affil{$^3$Geneva Observatory, University of Geneva, Maillettes 51, 1290 Sauverny, Switzerland}}

\begin{abstract}
If a massive star has lost significant mass during its red-supergiant stage,
it would return to blue region in the HR diagram and spend a part of 
the core-He burning stage as a blue supergiant 
having a luminosity to mass ratio ($L/M$) considerably 
larger than about $10^4$ (in solar units);
the duration depends on the degree of internal mixing and on the metallicity.
Then, various stellar pulsations are excited 
by enhanced $\kappa$-mechanism and strange  mode instability.
Assuming  these pulsations to be responsible for (at least some of) 
the quasi-periodic light 
and radial-velocity variations in $\alpha$ Cygni variables including 
luminous blue variables (LBVs; or S Dor variables), 
we can predict masses and surface compositions for these variables, and 
compare them with observed ones to constrain the evolutionary models.
We discuss radial pulsations excited in evolutionary models of an initial mass
of $40\,M_\odot$ with solar metallicity of $Z=0.014$, and compare them to
micro-variations in the two Galactic LBVs, HR Car and HD 160529. 
We have found that these stars should has lost more than half of the initial mass and 
their surface CNO abundances should be significantly modified from the original
ones showing partial H-burning products.
\end{abstract}

\section{Introduction}

\begin{figure}[t!]
\begin{center}
\epsfig{file=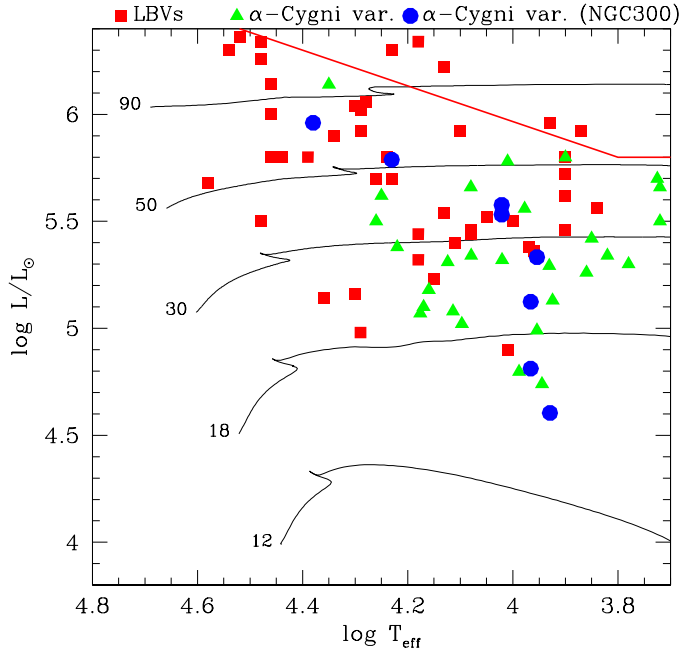,width=0.7\textwidth}
\end{center}
\caption{Approximate positions of $\alpha$ Cygni variables and LBV (S Dor) stars on the HR diagram. Straight lines in the upper-right part indicate the HD limit \citep{hum79}.  Observational parameters are adopted from \citet{vanG01} for LBV stars, 
\cite{vanL98} for other $\alpha$ Cygni variables in our Galaxy, LMC and SMC.
The parameters of  the variables in NGC 300  are adopted from \cite{bre04} and
\cite{kud08}.
}
\label{fig:hrd}
\end{figure}

Many luminous blue supergiants, including
Luminous Blue Variable (LBV, or S Dor variable) stars, show 
semi-periodic micro-variations in brightness and radial velocities, sometimes
superposed on long-timescale variations \citep[S Dor phases; e.g.,][]{lam98}.
The micro-variations are attributed to stellar pulsations, and these variables are called $\alpha$ Cygni variables.  
They  are known to be present in our Galaxy, Magellanic clouds \citep[e.g.,][]{vanL98}, 
and even in the nearby galaxy NGC 300 \citep{bre04}.
Fig.\,\ref{fig:hrd} shows approximate positions   
of some of the $\alpha$ Cygni  variables on the HR diagram. 
They are broadly distributed in the B-A supergiant region of 
$L \gtrsim 10^5\,L_\odot$,
in stark contrast to the distributions of classical variables confined to narrow
strips bounded by $T_{\rm eff}$s.
The latter property comes from the weakly nonadiabatic pulsations excited by
$\kappa$-mechanisms  around opacity peaks \citep[e.g.,][]{cox80},
while the distribution of the $\alpha$ Cygni variables indicate the semi-periodic
micro variations  to be extremely nonadiabatic pulsations
of stars with very high luminosity to mass ratios $(L/M\gtrsim 10^4\,L_\odot/M_\odot)$
involving strange modes \citep[e.g.,][]{sai09,sai13}.
For the evolution models of massive stars evolving toward the red-supergiant for
the first time, however,  such a high $L/M$ ratio occurs only for initial masses larger
than $\sim60\,M_\odot$ and hence the excitation of radial pulsations is expected
only for luminosities of $\log L/L_\odot\gtrsim 5.8$ \citep{kir93,sai11}; i.e., 
these models fail to explain the distribution of the $\alpha$ Cygni variables, which
extends down to $\log L/L_\odot \approx 4.6$ (Fig.\,\ref{fig:hrd}).

Recent evolution models including rotational mixing and updated wind mass-loss
rates \citep{eks12}, however, indicate that stars with masses larger than 
$\sim20\,M_\odot$ return to blue supergiant region
(blue loop) after considerable mass is lost during the red-supergiant stage 
\citep[the mass limit can be as low as $\sim12\,M_\odot$ if the mass-loss rates in the red-supergiant stage is enhanced greatly;][]{geo12}.
Significant mass loss makes the $L/M$ ratio on the blue loop 
much higher than that during the first crossing, so that  
radial/nonradial pulsations can be excited if   
$\log L/L_\odot\gtrsim4.7$ \citep{sai13}, which is roughly consistent with the distribution
of the $\alpha$ Cygni variables on the HR diagram.
Comparing with relatively less luminous $\alpha$ Cygni variables, 
\cite{sai13} found  that the excited periods are roughly consistent with 
most of the observed period ranges, although the predicted surface N/C and N/O
ratios seem too high at least for Rigel and Deneb.
 
In this paper, we discuss radial pulsations of slightly more luminous $\alpha$ Cygni
variables with S Dor type long-timescale variations (i.e., LBVs).

\section{Evolution of massive stars and the excitation of radial pulsations in LBVs}

\begin{figure}[h!]
\begin{center}
\epsfig{file=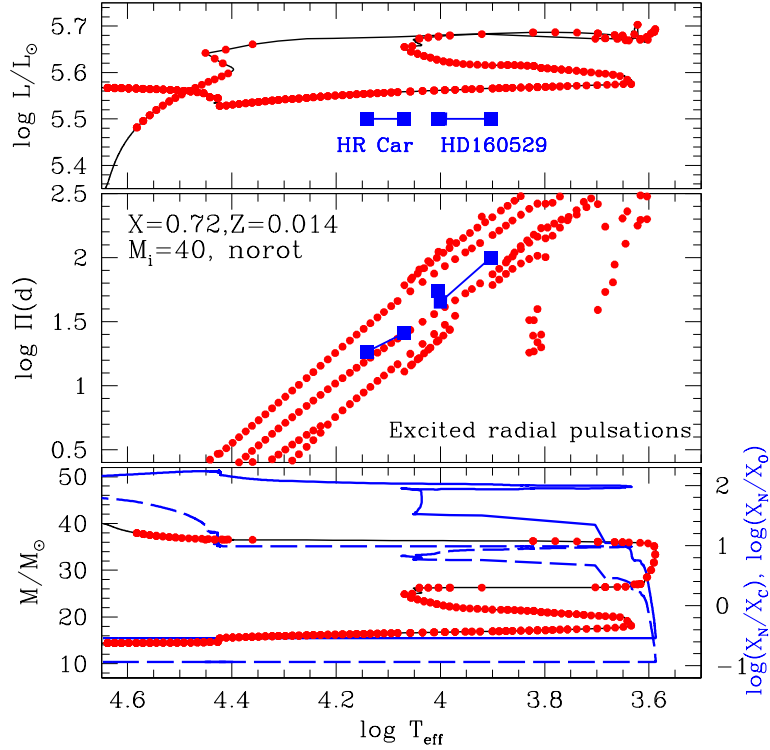,width=0.8\textwidth}
\end{center}
\caption{
{\bf Top panel:} Evolutionary track on the HR diagram  of a non-rotating model 
with an initial mass of $40M_\odot$, in which a core overshooting of $0.1$ 
pressure scale height is included, and the wind mass loss is included in the
same way described in \cite{eks12}. Approximate  positions of two galactic
LBVs, HR Car and HD 160529, are plotted, where parameters are adopted
from \cite{lam98} and \cite{sta03}. The effective temperature ranges connected
with horizontal lines indicate approximate ranges of variations during long-timescale
S Dor phases. Small red dots along the evolutionary track
indicate models in which at least one radial pulsation is excited.
{\bf Middle panel:} Periods ($\Pi$) of excited radial pulsations as function of
$T_{\rm eff}$ are compared with period ranges of  the micro-variations of the two LBVs 
shown in the top panel.  Periods are adopted from \cite{lam98} and \cite{sta03}.
{\bf Bottom panel:} The variations of total mass (black line) and of surface
abundance ratios, $\log (X_{\rm N}/X_{\rm C})$ (blue solid line) and
$\log(X_{\rm N}/X_{\rm O})$ (blue dashed line). 
The surface abundance ratios generally increase with evolution so that  
the post RSG phase corresponds to the upper part of the curves 
(see the blue scale on the right).
The red dots have the same
meaning as in the top panel.  They indicate pulsations are excited in blue 
supergiants only after the mass is decreased considerably.
}
\label{fig:lbvs}
\end{figure}

One useful property of the $\alpha$ Cygni type variations in LBVs is
that the period changes depending on the long-timescale S Dor phase; 
generally, the period is longer in visually brighter (i.e., cooler ) 
phases as discussed in \cite{lam98}; the bolometric luminosity is thought 
to be approximately constant during the S Dor phase. 
This phenomenon is consistent with interpreting the micro-variations
by stellar pulsation, of which period gets longer when the radius increases. 

Among the three Galactic LBVs discussed in \cite{lam98} we select two 
LBVs, HR Car and HD 160529, which have luminosities of
$\log L/L_\odot \approx 5.5$ comparable with luminosities during
the blue-loop evolution of our $M_{\rm i}=40\,M_\odot$ ($M_{\rm i}=$ initial mass)
model as shown in the top panel of  Fig.\,\ref{fig:lbvs}.
The ranges of $\log T_{\rm eff}$ connected with horizontal lines indicate
the ranges of temperature variations during long-timescale S Dor phases.

As we can see in Fig.\,\ref{fig:lbvs}, during the evolution toward the red-supergiant
region after the main-sequence evolution (1st crossing), no radial pulsations
are excited in the blue-supergiant region ($4.35 \gtrsim \log T_{\rm eff} \gtrsim 3.85$).
The model returns to blue during core He burning after  losing considerable mass
in the red-supergiant stage (Fig.\,\ref{fig:lbvs} bottom panel), and then
radial pulsations are excited. 
During the final blueward evolution, the model has a luminosity of
$\log L/L_\odot \approx 5.55$ comparable with the two LBVs.
In this phase of evolution, three radial pulsation modes are excited, and
their periods and the dependences on the effective temperature
are roughly consistent with the observed periods and their variations during
the S Dor phases of HR Car and HD 160529 (middle panel of Fig.\,\ref{fig:lbvs}). 
This indicates that these two LBVs would have already lost significant mass
from the initial one ($40 \rightarrow  17\,M_\odot$ in our model).  

Because of the large mass loss, the surface composition is modified, 
as layers where partial H-burning through CNO cycle occurred are progressively uncovered (bottom panel of Fig.\,\ref{fig:lbvs}). Since this model does not include rotational mixing, and since this model does not reach very low $T_{\rm eff}$ where very deep convective zone can develop, the major driver of the evolution of the surface chemical composition is the mass loss. As shown in Fig.\,\ref{fig:lbvs}, there are three major mass loss episodes, two at a $T_{\rm eff} \sim3.6$, and one at a $T_{\rm eff} \sim4.1$ (corresponding to the temperature of the bi-stability jump in the stellar wind where a sharp transition occurs in the mass-loss rates \citep{vin00}).
Most of the modification of the surface chemical composition occurs during the first strong mass-loss episode, where the star loses $\sim10\,M_\odot$, before the blue loop starts. The second mass-loss episode ($\sim5\,M_\odot$ lost) changes again slightly the surface composition at the tip of the loop. 
The last mass-loss episode, in the red part of the HRD, allows the star to evolve definitively towards the blue side of the HRD, but does not change the surface composition any more.

Our model predicts $X_{\rm H}=0.41$, $X_{\rm N}/X_{\rm C}=1.1\times10^2$,
and $X_{\rm N}/X_{\rm O}=9.7$ for the two LBVs, where $X_{\rm i}$ means the 
mass fraction of element i on the stellar surface. 
The surface hydrogen abundance corresponds to a number ratio of He/H$\approx0.35$,
which is comparable to the photospheric value of  P Cyg (another Galactic LBV), 0.40, 
listed in \cite{lam01}; P Cyg has a luminosity $\log L/L_\odot =5.8$
comparable to our model. 
Our surface $X_{\rm N}/X_{\rm O}$ ratio corresponds to a number ratio of  about 11,
which is higher than the ratios ($\le 6\pm 2$) in the nebulae around LBV stars 
discussed in \cite{lam01}. 
The difference might not be serious, because the ratio in nebulae should be 
considered as a lower bound for the photospheric value, and the observed
values might still be affected by considerable uncertainties.  
In addition, our present model does not include rotational mixing, which might change 
considerably the surface CNO ratios in the LBV stage. 

\section{Property of  very nonadiabatic pulsations of LBVs}
\begin{figure}[t!]
\hspace{-0.05\textwidth}
\epsfig{file=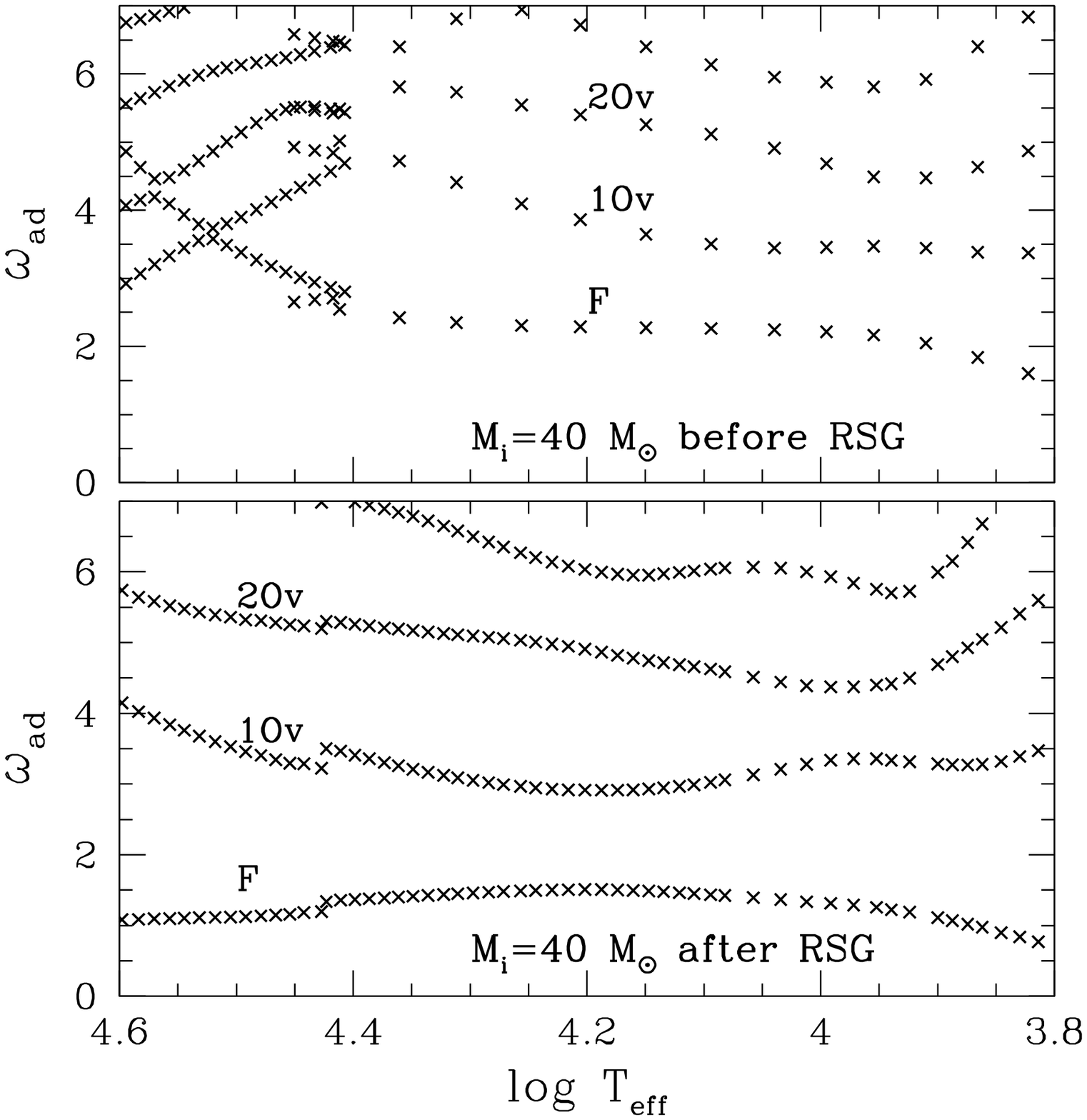,width=0.55\textwidth}
\hspace{-0.05\textwidth}
\epsfig{file=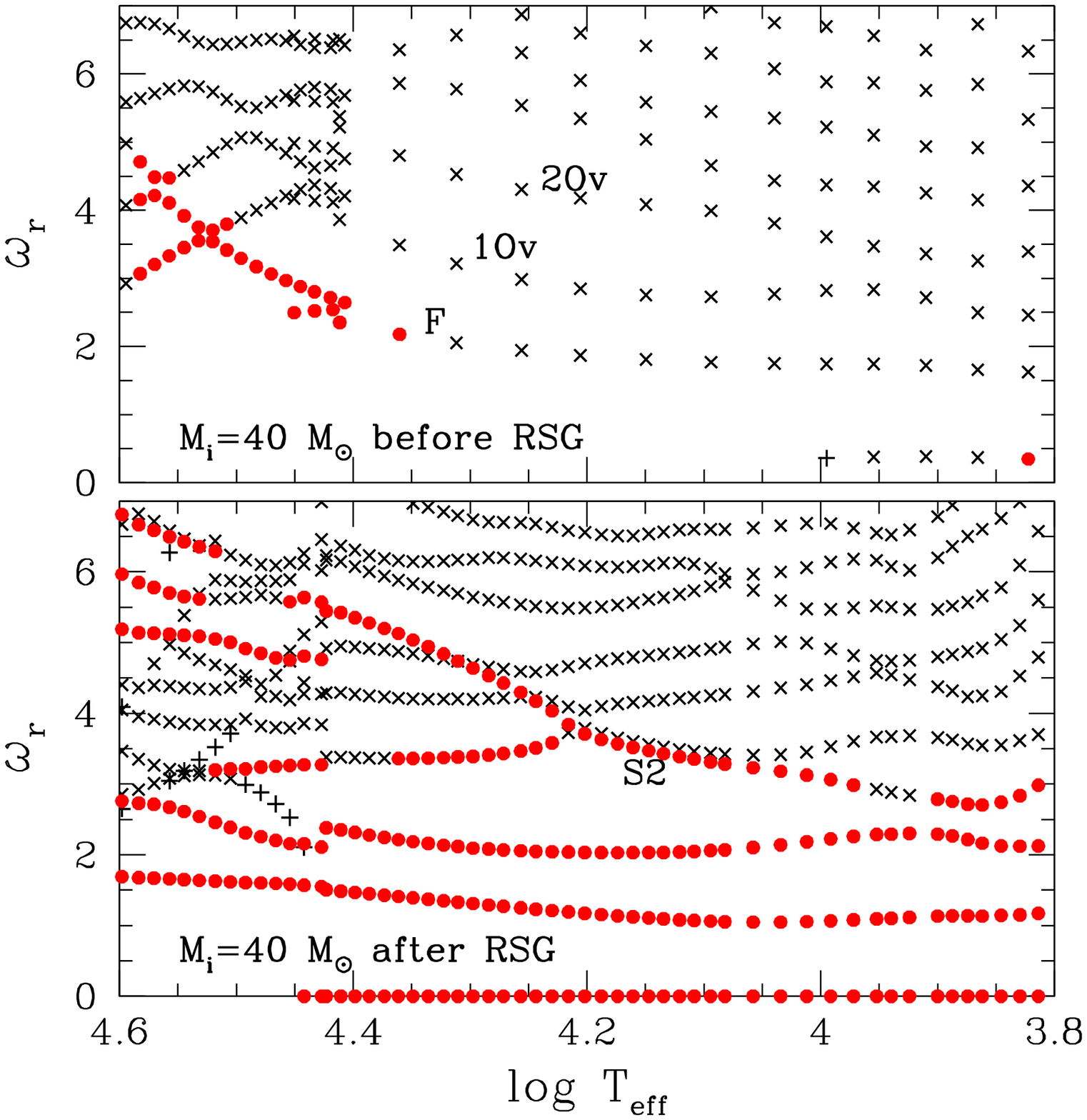,width=0.55\textwidth}
\caption{
Pulsation frequencies normalized by $\sqrt{GM/R^3}$ are plotted as a function
of effective temperature for models in the main-sequence stage and evolving to 
red-supergiant stage for 
the first time (top panels) and for models evolving blueward for the 2nd time
(4th crossing; bottom panels). The left panels show adiabatic frequencies and
the right panels show  non-adiabatic ones (the real part of eigenfrequencies; 
$\omega_{\rm r}$).  
Red dots indicate excited modes and crosses indicate damped modes.
In the left panels for adiabatic pulsations and in the top-right panel, fundamental
(F), first overtone (1Ov) and second overtone (2Ov) modes are indicated, 
but not in the bottom-right panel, because in the 4th crossing models, 
pulsations are so non-adiabatic that mode identifications are not possible. 
The symbol `S2' indicates genuine strange modes which exist even in the 
limit of diminishing thermal time. Note that along the sequence of excited
modes, damped modes tend to have similar frequencies, which form approximately
complex-conjugate pairs (see the top-left panel of Fig.\,\ref{fig:eigf}). 
}
\label{fig:omega}
\end{figure}

Fig.\,\ref{fig:omega} shows pulsation frequencies of several modes normalized
by $\sqrt{GM/R^3}$ as a function of effective temperature along the
evolutionary track of $M_{\rm i}=40\,M_\odot$ model 
during main-sequence and the post-main sequence 
(1st crossing) evolution in the top panels, and during the final blueward evolution
in the bottom panels.  
The left and right panels show adiabatic and nonadiabatic frequencies,
respectively.
Adiabatic pulsation modes are well ordered from lowest to higher frequencies as
fundamental (F), 1st overtone (1Ov), 2nd overtone (2Ov), etc even in the advanced
evolution stage.
However, the ordering is destroyed by strong nonadiabatic effects as in the model on the
final blueward evolution (right-bottom panel of Fig.\,\ref{fig:omega}), in which
it is not possible to assign the correspondence between  adiabatic and 
nonadiabatic modes.

Nonadiabatic (real parts of ) frequencies of excited modes are plotted by red dots
(damped modes by crosses).  
The right-bottom panel of Fig.\,\ref{fig:omega} indicates that during the
blue-supergiant phase on the blue loop three pulsation modes and one monotonic
mode ($\omega_{\rm r}=0$) are excited. 
Note that the sequence denoted as S2 consists, in most part, two modes having
similar pulsations frequencies ($\omega_{\rm r}$) but one is excited and the other
damped. These two modes approximately form a complex-conjugate pair; 
this is the property of genuine strange-mode instability in the extremely nonadiabatic
condition with diminishing thermal time \citep[NAR condition;][]{gau90}.

\begin{figure}[t!]
\hspace{-0.04\textwidth}
\epsfig{file=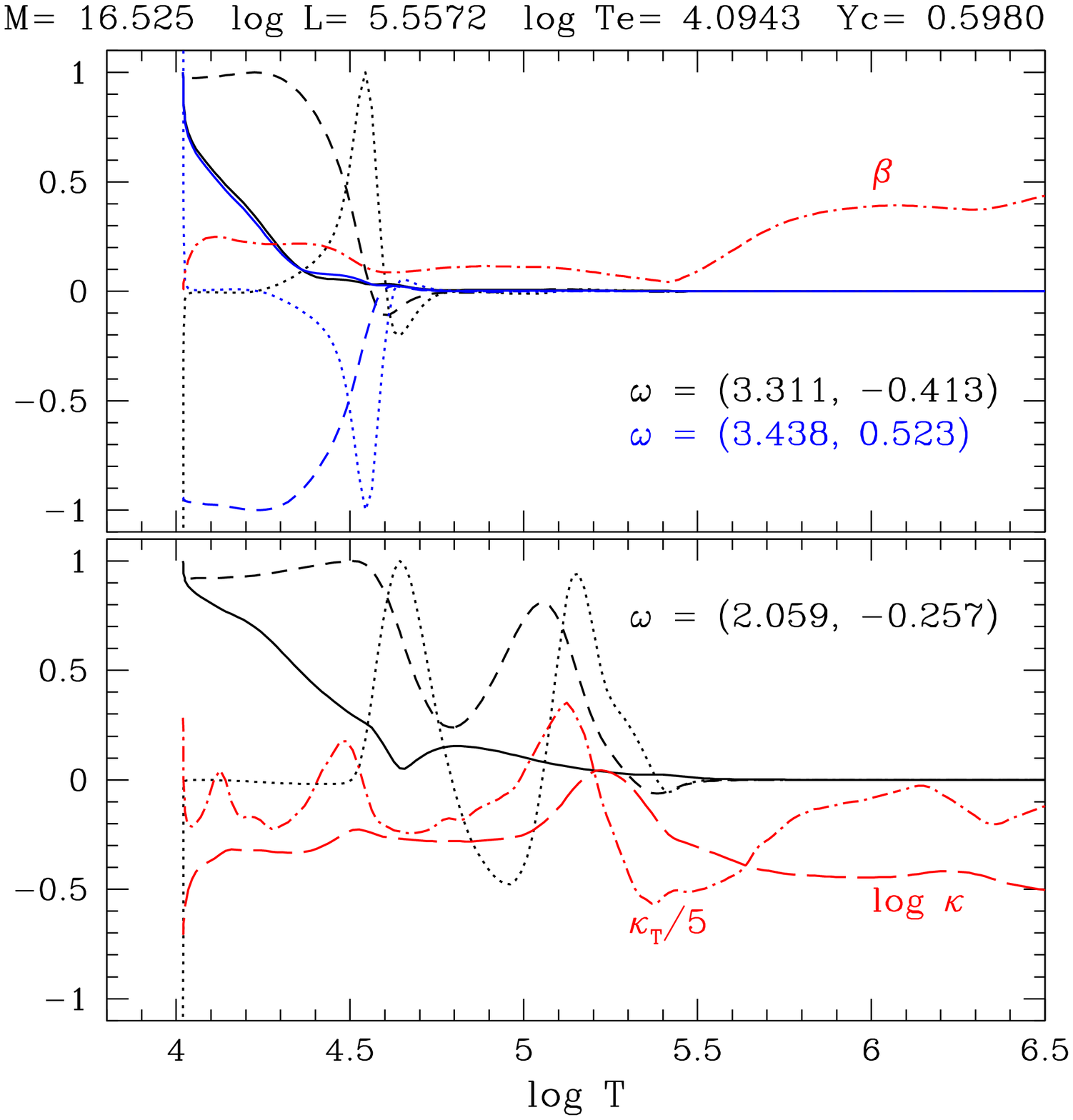,width=0.55\textwidth}
\hspace{-0.06\textwidth}
\epsfig{file=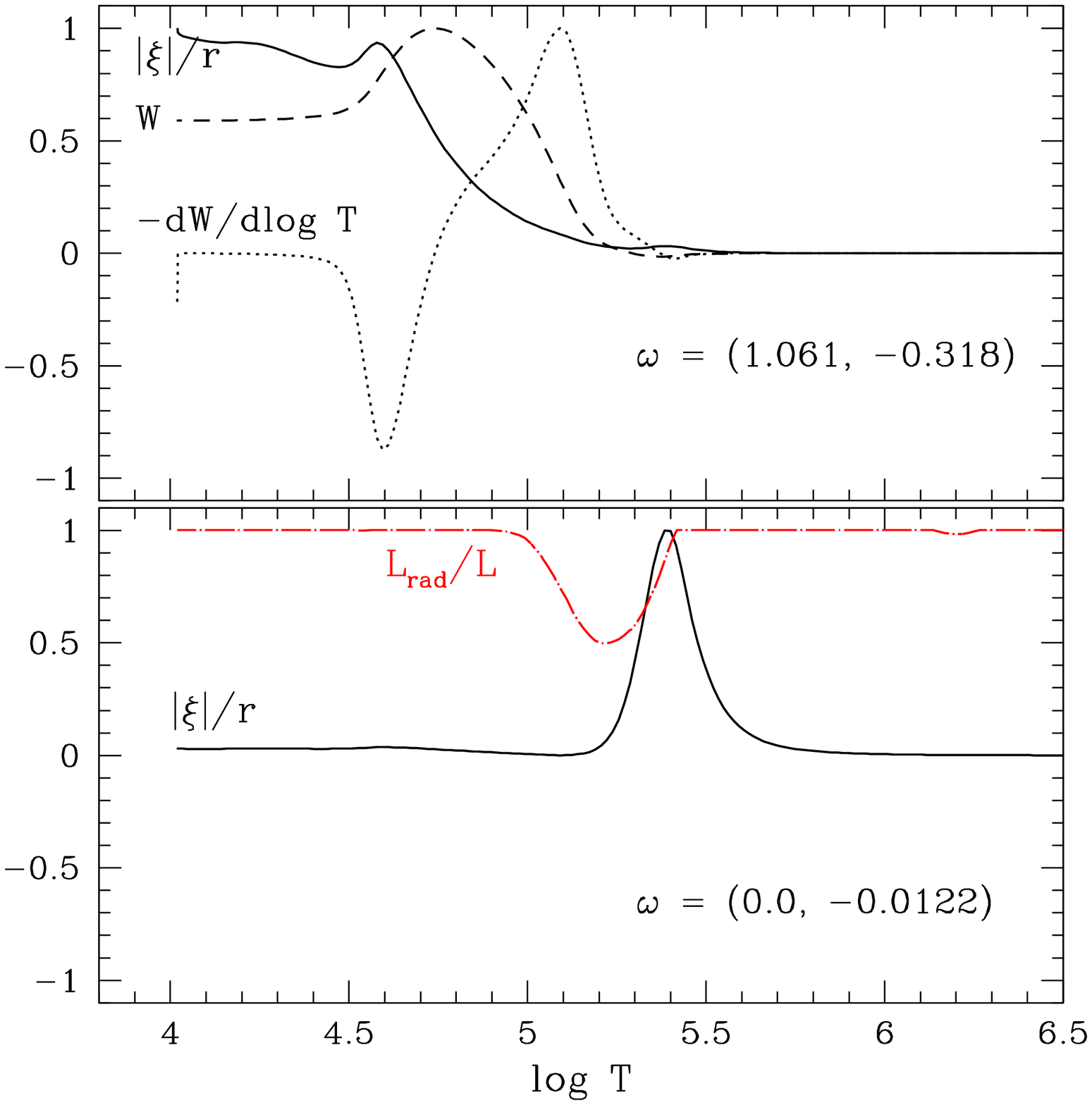,width=0.55\textwidth}
\caption{
Profiles of displacements $|\xi|/r$ (solid lines), and work, $W(r)$ (dashed lines), 
and differential work, $-dW/d\log T$ (dotted lines), 
are shown for the four lowest frequency modes 
in the model evolving blueward (right-bottom panel of Fig.\,\ref{fig:omega})
at $\log T_{\rm eff}=4.09$.  
Layers with $-dW/d\log T >0$ drive the pulsation, and the pulsation is excited 
if the work $W$ is positive at the surface
\citep[see][for the definition of $W$]{sai13}. 
The blue lines in the top-left panel are for a (damped) mode nearly
complex-conjugate of the excited one shown by black lines.
Note that the blue line for the displacement $|\xi|/r$ is nearly superposed on the 
black line, and the works and the differential works for the two modes 
are nearly mirror-symmetric.
Red lines show profiles of $\beta=P_{\rm gas}/P$, opacity $\kappa$ and 
its derivative with respect to the temperature, $\kappa_T\equiv d\ln\kappa/d\ln T$,
and the ratio of the radiative to total luminosity $L_{\rm rad}/L$.
}
\label{fig:eigf}
\end{figure}

Fig.\,\ref{fig:eigf} shows some of the properties of the excited modes and the
damped complex-conjugate mode in the model at $\log T_{\rm eff}=4.094$ on
the final blueward track (the mass is  reduced to $16.5\,M_\odot$).
The top-left panel is for the S2 pair modes; black lines for excited modes
and blue lines for damped modes. The listed complex frequencies indicate
that these modes are indeed nearly complex-conjugate to each other.
Amplitude profiles of these modes are nearly identical and confined to the second
He ionization zone and above where the radiation pressure is dominant 
($\beta=P_{\rm gas}/P\ll 1$) and the thermal time is very small; 
this explains why these modes have extreme nonadiabatic strange-mode property.

The mode shown in the left-bottom panel is excited by enhanced $\kappa$-mechanism
at He II ionization zone ($\log T\approx 4.6$) and at the Fe opacity bump
($\log T\approx5.2$), and the mode in the right-top panel is excited at 
the Fe opacity bump.

The amplitude of the mode shown in the right-bottom panel of 
Fig.\,\ref{fig:eigf}  monotonically grows; i.e.
a small deviation from the equilibrium state grows monotonically, 
although the growth time  is some thirty times longer than 
that of the excited pulsation modes.
\citep[Such monotonically growing modes were also found in first crossing models of $M_{\rm i} \ge 60\,M_\odot$;][]{sai11}
This mode has a large amplitude around $\log T\sim 5.4$ slightly deeper than 
layers where the Fe-opacity bump resides.
The significance of the monotonic modes in massive star phenomena is not clear.

\section{Conclusion}
We discussed radial pulsations excited in evolutionary models of  
$M_{\rm i} =40\,M_\odot$ and found that the pulsations are excited during the
blueward evolution after the mass has been reduced to $\sim17\,M_\odot$. 
They seem consistent with the periods and their variations of the micro variations
observed in the two Galactic LBVs, HR Car and HD 160529.
This indicates that these stars should have already lost significant mass, and 
the surface compositions should have been modified significantly from the
original ones showing partially processed H-burning products.

Finally, we note that pulsations 
might be able to significantly affect mass loss rates in LBV (and pre-LBV) stars.
For example, the nonlinear pulsation analysis by \cite{gla99} for a supergiant model 
of  $64\,M_\odot$ ($M_{\rm i} = 120\,M_\odot$) indicates 
that the photospheric velocities 
of pulsation can reach to the escape velocity.  
Observationally, on the other hand, 
\cite{aer10} found that the mass loss rates of the luminous
($\log L/L_\odot \approx6.1$) blue supergiant HD 50064 change on a time scale
similar to the period of photospheric variations, indicating connection between
pulsation and mass loss.
Further theoretical and observational investigations on the interaction 
between pulsation and mass loss in massive stars would be important.

\acknowledgements 
CG acknowledges support from EU-FP7-ERC-2012-St Grant 306901.

\bibliography{ref_hs}

\end{document}